\documentclass[a4paper, aps, pra, preprintnumbers, twocolumn]{revtex4} 
\usepackage{graphics, graphicx, epsfig}
\usepackage[centertags]{amsmath}
\usepackage{amsfonts}
\usepackage{amssymb}
\usepackage[cp1251]{inputenc}
\usepackage[english]{babel}
\usepackage{amsthm, color}

\begin{document}

\def\BE{\begin{equation}}
\def\EE{\end{equation}}
\def\BEA{\begin{eqnarray}}
\def\EEA{\end{eqnarray}}
\def\BY{\begin{eqnarray}}
\def\EY{\end{eqnarray}}

\def\L{\label}
\def\nn{\nonumber}
\def\ds{\displaystyle}
\def\o{\overline}

\def\({\left (}
\def\){\right )}
\def\[{\left [}
\def\]{\right]}
\def\<{\langle}
\def\>{\rangle}

\def\td{\tilde}
\def\pr{\prime}

\def\k{\mathbf{k}}
\def\q{\mathbf{q}}
\def\r{\vec{r}}
\def\ro{\vec{\rho}}
\def\a{\hat{a}}
\def\b{\hat{b}}
\def\c{\hat{c}}
\def\e{\hat{e}}
\def\hs{\hat{\sigma}}
\def\h{\hat}

%
\title{Two-photon bunching inside a quantum memory cell} \vspace{1cm}
\author{ A.S.~Losev$^1$, {T.Yu.~Golubeva$^1$},  {A.D.~Manukhova$^2$}, {Yu.M.~Golubev$^1$}, }
\address{$^1$Saint Petersburg State University, 7/9 Universitetskaya nab., St. Petersburg, 199034, Russia \\
$^2$Department of Optics, Palack{\'y} University, 17. Listopadu 12, Olomouc, 771 46, Czech Republic}
\begin{abstract}

We demonstrate the effect of interference, similar to the Hong-Ou-Mandel effect, for two single-photon pulses by writing them sequentially inside a quantum memory cell in a tripod configuration.

\end{abstract}
\maketitle
%
\section{Introduction}\L{I}

Synchronization of quantum single-photon pulses is a complex issue motivated by the use of single-photon quantum fields as a resource in problems of quantum communication and quantum computing ~\cite{Felinto-2006, Knill-2001, Nagao-2019, Carlos-2011}. 
Progress in creation of pure single-photon quantum states achieved in recent years~\cite{Tanida-2012, He-2013, Gazzano-2013, Flagg-2013, Bernien-2012, Chaneliere-2007, Kurz-2013, Maunz-2007, Beugnon-2006}, 
has been accompanied by new challenges: 
single-photon fields should be fully time-controlled for their successful application in the protocols of quantum repeaters~\cite{Sangouard-2011} or quantum computing networks~\cite{Felinto-2006}. 
The randomness of the photon creation process is a fundamental obstacle to the scalability of quantum networks.

Quantum memory cells became one of the natural proposals for synchronizing the processes of single-photon generation~\cite{Felinto-2006, Yuan-2007}. 
As devices enabling one to write, store, and readout a quantum signal on demand, such cells are the desired control element that allows one to overcome the random nature of the photon generation process.
 
One of the first successful experiments on the synchronization of single-photon pulses was performed by the group of Akira Furusawa~\cite{Furusawa}. 
However, these experiments were based on resonant photon delay lines and not on-demand quantum memory cells. 
Note that on-demand memory cells are devices operating on the principle of transferring the quantum state of the field to the long-lived collective coherence of the medium and then, after the storage, back to the field. 
The resonant delay line works differently; it assumes the nonlinear conversion and creation of a photon pair inside each of the devices, 
 so one photon from the pair serves as a herald of the other~\cite{Lvov}.

Experimental research on two-photon interference of the weak coherent laser pulses stored in two spatially separated solid-state memory cells in the AFC(atomic frequency comb) scheme was carried out in~\cite{Tittel}. 
Similar studies, but using cells with rubidium vapor in the EIT(electromagnetically induced transparency) mode at room temperature, were performed in~\cite{Scriminich}.

It should be noted that memory cells turned out to be very sensitive even to small differences in the properties of ensembles and conditions for their interaction with quantum fields~\cite{Tikhonov-2015}. 
So the creation of the two identical cells for synchronizing single-photon fields required a virtuoso experimental technique.
 
In the works cited above, the photon bunching effect and the Hong-Ou-Mandel state~\cite{Hong-1987} were considered only as a verification method, 
which served as an indicator of good synchronization of the wave trains. 
At the same time, two-photon states obtained as a result of Hong-Ou-Mandel interference are of independent interest. 
They can be considered as NOON states~\cite{NOON}, 
and in this way, a sufficiently reliable scheme for obtaining such states can be useful in quantum metrology or in sensitivity enhancement problems, where NOON states are usually used.
 
In our work, we propose a way to synchronize two single-photon fields, requiring the only one memory cell in a tripod configuration~\cite{Hamedi-2019, Losev-2017, Losev-2016, Ruseckas-2011, Paspalakis-2002}, 
which can significantly simplify the technical side of the experiment.

We consider the scheme when the synchronization (and destructive interference) of photons occurs not after the memory cells, but directly inside the cell. 
In fact, since the Hamiltonian of the interaction of light and matter in the memory cell is similar to the Hamiltonian of the beam splitter~\cite{Losev-2016, Losev-2017}, 
one can expect to see an analog of the beam splitter effect for two single-photon fields that are sequentially sent to the cell inputs.

Here, we discuss the conditions for observing the Hong-Ou-Mandel effect when two single-photon pulses are written sequentially in a quantum memory cell in a tripod configuration.
We indicate the conditions for controlling the process of writing and readout by the classical fields that provide the interaction of a beam splitter type 
and observing the effect of photon bunching at the output of a memory cell.

\section{Hong-Ou-Mandel effect at the output of the broadband tripod memory}
%
\subsection{Mixing of single-photon pulses inside a two-channel memory cell}

As is known, the Hong-Ou-Mandel effect occurs when two identical statistically independent single-photon pulses are mixed on a symmetric beam splitter, 
so that at the output of the beam splitter the photon bunching effect arises and the output state takes the form:
\BE |out\> = \frac1{\sqrt2} \( |2\>_{1}|0\>_{2} -
|0\>_{1}|2\>_{2} \), \L{1} \EE
where the indexes $ {1,2} $ number the beam splitter channels.

That means, that at the output both photons are localized in the same pulse, although at the input both pulses were single-photon pulses.

To observe this effect, it is crucial to achieve a coordinated (simultaneous) arrival of both initial pulses to the beam splitter.
Thus, it is an excellent test for strict synchronization of two single-photon pulses.
In practice, it is impossible to synchronize two independent pulses without special efforts, since the initial pulse creation is usually of a statistical nature.
Therefore, we can only talk about the probability of a photon arriving at a beam splitter at a given time.

As we know, quantum memory cells and beam splitters quite often "work" \; similarly~\cite{Losev-2016}.
Here we consider the case where a beam splitter is replaced by a tripod (two-channel) memory cell~\cite{Losev-2017}.
The thought experiment in this case is as follows.
We assume that the input signal pulses $\hat a_1 (t, z) $ and $ \hat a_2 (t, z) $ are created in two statistically independent processes
and each goes into the memory cell via its own channel (tripod configuration of the medium).
Let the first single-photon pulse appear at the input of the memory cell.
We will write it in one of the two free channels.
Then we wait for the arrival of the second single-photon pulse and write it in the other free channel.
As a result, both spin waves of the tripod memory are excited.
We assume these excitations to be independent, identical, and sufficiently long-lived.
Furthermore, we will see that during the readout from both channels of the quantum memory, photon bunching will occur, so that both photons will be read together into one of the channels,
and the radiation wave function at the output of the cell takes the form Eq.~\ref{1} (so it appears to be in the HOM state).
This means that inside the cell the quantum states of the fields interfere and we observe the Hong-Ou-Mandel effect.

\begin{figure}[h]
\centering
\includegraphics[height=40mm]{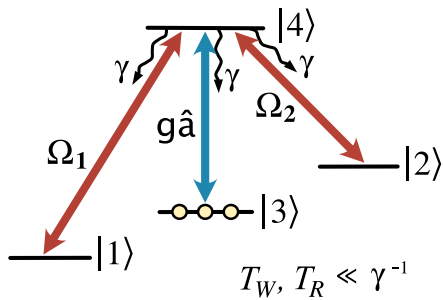}
\caption{
Tripod configuration of the atomic levels and transitions:
$ \Omega_1 $, $ \Omega_2 $ are the Rabi frequencies of the driving fields,
$ g $ is the coupling constant,
$ \a $ is the quantized amplitude of the signal field,
$ \gamma $ is the relaxation rate from the excited state $ | 4 \> $,
$ T_W, T_R $ are the writing and readout times.
Transitions between the three lower levels are dipole forbidden.
Initially only the third level is populated.
} \L{Fig1}
\end{figure}

Two classical rectangular-shaped driving pulses with Rabi frequencies $ \Omega_1 $ and $ \Omega_2 $ participate in the writing and readout processes,
acting resonantly on the transitions $ | 1 \>  \! - \!  | 4 \> $ and $ | 2 \>  \! - \!  | 4 \> $, respectively.
Two quantum signal pulses, which are sequentially written in the memory cell, resonantly interacting with the ensemble at the transition $ | 3 \>  \! - \!  | 4 \> $
(i.e., the carrier frequencies of the pulses coincide with the corresponding frequencies of the atomic transitions).
In this problem, we use the fast quantum memory protocol~\cite{Golubeva-2012}, which means 
that the writing time $ T_W $ and reading time $T_R$ (assuming $ T_R = T_W $), and therefore the durations of the written and read pulses which coincide with them,
are much less than the lifetime of atoms in the excited state $ | 4 \> $: $ T_W, T_R \ll \gamma^{- 1} $,
where $ \gamma $ is the atomic relaxation rate of the atoms from the upper state $ | 4 \> $ to any of the lower ones (Fig.~\ref{Fig1}).
 
Note that the effect under discussion is not limited only by this memory protocol which we have chosen as an example to be able to perform some specific calculation and demonstrate the effect.

Writing to the memory cell and the readout from it are carried out by choosing the driving fields.
We have two driving fields.
One, with Rabi frequency $ \Omega_1 $, acts in resonance with the atomic transition $ | 4 \rangle - | 1 \rangle $.
Another with frequency $ \Omega_2 $ acts on the transition $ | 4 \rangle - | 2 \rangle $.
With the joint action of the driving and signal, the signal fields (or some parts of them) are written into the memory cell.
In our consideration,  the spin wave with the Heisenberg amplitude $ \hat b_1 (z) $ is excited first in the mode $ \Omega_1 = \Omega, \; \Omega_2 = 0 $.
Then, in the mode $ \Omega_2 = \Omega, \; \Omega_1 = 0 $, a spin wave with an amplitude $ \hat b_2 (z) $ is excited.

We will assume that readout is also performed in two steps.
At the first stage, which is determined by the set $ \Omega_1 = \Omega_2 = \Omega $, there is partial readout from both spin channels.
As a result, at the output of the memory cell, at the point $ z = L $, a field with the amplitude $ \hat a _ {+} (t, z = L) $ will appear, which we denote below as $ \hat a _ {+} (t) $.
According to~\cite{Losev-2016}
\BY
&&\hat a_{+}(t)=  \frac1{\sqrt{2}} \int^{T_W\!\!\!\!\!}_0 dt^\prime
\;G(T_W - t,T_W -t^\prime) \times  \nn\\
&&\[\hat a_{1}(t^\prime) + \hat a_{2}( t^\prime)
\]+\h{v}_{+}(t),\ \ \hat a_{1,2}( t)=\hat a_{1,2}( t,z\!=\!0).\qquad \L{2}
\EY
At the end of the first stage of the readout, we proceed to the second stage determined by the set of the driving fields $ \Omega_1 = - \Omega_2 = \Omega $.

\BY
&&\hat a_{-}(t)=  \frac1{\sqrt{2}} \int^{T_W\!\!\!\!\!}_0 dt^\prime
\;G(T_W - t,T_W -t^\prime)\times  \nn\\
&&\[\hat a_{1}(t^\prime) - \hat a_{2}( t^\prime)
\]+\h{v}_{-}(t),\ \ \hat a_{-}( t)=\hat a_{-}( t,z=L).\qquad \L{3}
\EY
From here on, the amplitude is expressed in terms of the dimensionless coordinate $ z $ and dimensionless time $ t $, which are introduced according to the relations
\BE
\Omega t\to \! t, \ \ \Omega T_W \! \to T_W, \ \
{2g^2Nz}/{\Omega}\to \! z, \ \ {2g^2NL}/{\Omega}\to \! L,\nn
\EE
where $ g $ is the coupling constant in the interaction of quantum fields with the medium and $ N $ is the linear concentration of the atomic ensemble.

As a result of two readout steps, two pulses appear at the output of the memory cell at the point $ z = L $.
As we will see later, these pulses are in the Hong-Ou-Mandel state, in contrast to the input fields at the point $ z = 0 $, where each of the two fields was in a single-photon state.

The amplitudes $ \hat v_\pm $ are related to the vacuum components of the field.
They are introduced in order to ensure preservation of the commutation relations for the field amplitudes, and their commutation relations can be written as
\BE
\[\hat v_{\pm}(t), \hat v^\dag_{\pm}(t^\prime)\] = \delta(t - t^\prime) -\!\! \int_0^{T_W\!\!\!\!\!}dt_1\,G(t,t_1)\,G(t^\prime,t_1)
.\L{4}
\EE
In the case of the ideal quantum memory, where the kernel $ G (t, t^\prime) $ is represented as $ \delta(t-t ^\prime) $,  we can get
\BE
\hat a_{\pm}(t)=   \[\hat a_{1}(t) \pm \hat a_{2}( t)
\]/\sqrt{2}.\L{5}
\EE
instead of Eqs.~(\ref{2}) - (\ref{3}).
Note that the role of vacuum components in this case turns out to be zero, since $[\hat v_{\pm}(t),\hat v^\dag_{\pm}(t^\prime)]=~0$ according to~Eq.~(\ref{4}).

\subsection{Hong-Ou-Mandel effect via the two-channel memory} 

The equalities  (\ref{2})-(\ref{3})  depend on the kernel $G(t,t^\prime)$,
which directly connects the signal amplitudes at the output of the memory cell $\hat a_\pm(t)$ with the amplitudes at the input $\hat a_{1,2}(t)$.
According to \cite{Tikhonov-2015EurPhys, Losev-2017} for broadband resonant memory $G(t,t^\prime)=G(t^\prime,t)$.
Therefore we can write the Schmidt decomposition as
\BE
G(t,t^\prime) = \sum_{i=1}^\infty \sqrt{\lambda_i}\;
\psi^\ast_i(t) \psi_i(t^\prime). \L{6}
\EE
Here, $ \lambda_i $ is the set of eigenvalues of the matrix $ G (t, t^\prime) $ and $ \psi_i (t) $ is the complete orthonormal set of the eigenfunctions.
The amplitudes can be decomposed using this set,
\BE
\hat a_{1,2}(t)  = \sum_{i=1}^\infty \psi _i(t) \, \h{e}_{1 ,
2;i},
\quad\hat a_\pm(t)  = \sum_{i=1}^\infty \psi_i(t) \,
\h{e}_{\pm,i}. \L{7}
\EE
Here, $ \h{e} _i $ are discrete canonical operators for which the following hold
\BE
\h{e}_i = \int_0^{T_W\!\!\!\!\!} dt \, \psi_i^\ast(t)\a(t),\quad \[\h{e}_i, \h{e}_j^\dag \] = \delta_{ij} . \L{8}
\EE
The conditions of orthonormality and completeness for eigenfunctions are as follows
\BE
\int_0^{T_W\!\!\!\!\!}  dt\;\psi_i^\star(t) \psi_j(t) = \delta_{ij},\ \
\sum_{i=1}^\infty \psi_i^\star(t) \psi_i(t^\prime)
=\delta(t-t^\prime). \L{9}
\EE
Taking into account the Schmidt decomposition~(\ref{6})-(\ref{7}),  we obtain the following equalities
\BE
\h{e}_{\pm,i} =   \sqrt{\lambda_i}\;\frac1{\sqrt{2}} \(
\h{e}_{1,i} \pm \h{e}_{2,i} \) + \sqrt{1-\lambda_i} \,
\h{e}_{vac,i}. \L{10}
\EE
instead of Eqs.~(\ref{2})-(\ref{3}).

For the ideal memory considered above, all the eigenvalues $ \lambda_i $ are equal to one, so the Eq.~(\ref{10}) takes the form
\BY
&& \h{e}_{\pm,i} =    ( \h{e}_{1,i} \pm \h{e}_{2,i}
)/\sqrt{2}\L{11} .
\EY
Now we can write the wave functions of the field at the input and at the output of the memory cell
\BY
&|in\rangle = \prod_i |in\rangle_i , \qquad &|in\rangle_i =  \h e_{1,i}^\dag \h
e_{2,i}^\dag |0\rangle_i,\nonumber\\
&|out\rangle =\prod_i|out\rangle_i , \quad &|out\rangle_i
= \h e_{+,i}^\dag \h e_{-,i}^\dag|0\rangle_i \L{12} .
\EY
Hence we obtain that the two pulses of light at the output of the quantum memory turn out to be in the Hong-Ou-Mandel state:
\BE
|out\rangle_i =
\(|2\rangle_{1i}|0\rangle_{2i}-|2\rangle_{2i}|0\rangle_{1i}
\)/\sqrt 2\L{13} .
\EE
The result is the analogue of the Hong-Ou-Mandel effect.
This predicts that when two single-photon pulses are mixed in the memory cell,  the output photons definitely will  be in one of the two read pulses.

In \cite{Losev-2016}, we discussed a memory process, where only one eigenvalue $ \lambda_1 $ was noticeably different from zero: $ \lambda_1 \approx 1 $ and $ \lambda_{i> 1} \ll1 $.
Then, according to Eq.~(\ref{10}), only one field component with $i = 1$ will be in the Hong-Ou-Mandel state.

\section{Conclusion}

We demonstrated the two-photon interference inside a quantum memory cell when two single-photon pulses are written sequentially to it.

The construction in the language of the Schmidt mode given in the previous section allows us to state 
that in a multimode quantum memory the photon bunching effect, similar to the Hong-Ou-Mandel effect, appears only for the eigenmodes of the cell corresponding to eigenvalues close to unity.
It is these modes that are converted during the storage like mixing on a beam splitter.

It is interesting to note that photon interference is observed despite the fact that the readouts of the pulses are carried out at different points in time. 
That is, the temporal separation of pulses using a memory cell is similar to the spatial one using a beam splitter. 
In this sense, the resulting NOON state is similar to temporally separated and spatially bunched (TSSB) state \cite{Kim-2016}.

\section{Acknowledgment}

This work was supported by the Russian Foundation for Basic Research (Grant Nos 19-02-00204a and 18-02-00648a).
A.M. acknowledges the Project 20-16577S of the Czech Science Foundation and the project CZ.02.1.01/0.0/0.0/16\_026/0008460 of MEYS CR.


\end{document}